\documentclass[showpacs,preprintnumbers,
amsmath,amssymb,aps,prd,nofootinbib,superscriptaddress,
eqsecnum,12pt]{revtex4}
\usepackage{graphicx}
\draft

\makeatletter
\renewcommand{\p@subsection}{}
\makeatother

\renewcommand{\thesection}{\arabic{section}}

\renewcommand{\theequation}{\arabic{section}.\arabic{equation}}

\newcommand{\mbf}[1]{{\mbox{\boldmath $#1$}}}
\newcommand{\tr}{\textrm{tr}}

\newcommand{\chushi}[1]{ }

\begin{document}

\title{Masses of vector bosons in two-color dense QCD\\
based on the hidden local symmetry}

\author{Masayasu Harada, Chiho Nonaka, and Tetsuro Yamaoka\\
{ \small
\textit{Department of Physics, Nagoya University,
Nagoya, 464-8602, Japan}
}
}

\date{\today}

\begin{abstract}
We construct a low energy effective Lagrangian for the
two-color QCD including the ``vector'' bosons
(mesons with $J^P=1^-$ and diquark baryons with $J^P=1^+$)
in addition to the pseudo Nambu-Goldstone bosons
with a degenerate mass $M_\pi$
(mesons with $J^P=0^-$ and baryons with $J^P=0^+$)
based on the chiral symmetry breaking pattern of 
$SU(2N_f) \rightarrow Sp(2N_f)$
in the framework of the hidden local symmetry.
We investigate the dependence of the ``vector'' boson masses
on the baryon number density $\mu_B$.
We show that
the $\mu_B$-dependence
signals the phase transition of $U(1)_B$ breaking.
We find that it gives information about
mixing among ``vector'' bosons:
e.g.~the mass difference between $\rho$ and
$\omega$ mesons is proportional to the
mixing strength between the diquark baryon with $J^P=1^+$
and the anti-baryon.
We discuss the comparison with lattice data for two-color QCD
at finite density.
\end{abstract}

\pacs{11.30.Rd, 12.39.Fe, 05.70.Fh}

\maketitle

\section{Introduction}
\label{sec:intro}

Quantum chromodynamics (QCD)
shows various phases under extreme conditions.
At very high temperature and/or density,
the phase structure can be studied
with perturbative approaches.
However, it is difficult to study it
especially near the critical temperature
and/or density directly from QCD,
because of the strong coupling.
Lattice QCD simulation is one of powerful theoretical tools,
but it is not applicable in the finite density region
due to the sign problem~\cite{sign}.
The problem in simulations of real-life QCD 
at finite baryon density produces
interest in the two-color QCD with quarks in the fundamental 
representation~\cite{2-color-eff,KSTVZ},
the QCD with quarks in the adjoint representation~\cite{adjoint,KSTVZ},
and so on,
that are free from the sign problem.
In particular, two-color QCD has some interesting features
in the following respect:
Color-singlet baryons appear
together with ordinary mesons
as Nambu-Goldstone (NG) bosons associated with
the spontaneous breaking of the
chiral symmetry and their interactions are determined
uniquely by the low-energy theorem.
This allows us to construct low-energy effective theories
including the baryons as light degrees of freedom naturally,
and to investigate properties of hadrons
even at finite baryon density using them.
Furthermore, because there are
many studies of the lattice QCD simulations,
we can make a comparison of results from effective theories
with those from the lattice analyses~\cite{2-color-lat,MNN,HSS}.

Actually, in two-color QCD,
the phase structure at finite baryon number density $\mu_B$
was studied~\cite{KSTVZ} 
using the chiral Lagrangian including the
pseudo-NG bosons, having masses of $M_{\pi}$,
which are associated with the chiral symmetry breaking of
$SU(2N_f) \rightarrow Sp(2N_f)$.
It was shown that
the phase transition from the symmetric phase
of the baryon number $U(1)_B$ to the broken
phase takes place at $\mu_B = M_{\pi}$,
triggered by the condensation of the baryonic pseudo-NG bosons.
There, the density dependences of the masses of pseudo-NG bosons 
are also studied and it was found that a baryonic boson becomes the
massless NG boson in the $U(1)_B$ broken phase.

The density dependences of ``vector'' bosons
(In this paper we call ``vector'' bosons 
which consist of mesons with $J^P=1^-$
and diquark baryons with $J^P=1^+$.)
are studied 
in the lattice simulation~\cite{MNN,HSS}.
It was shown that
the mass of ``$\rho$'' meson decreases with $\mu_B$~\cite{MNN,HSS},
and that
the mass of anti-baryon with $J^P = 1^+$ increases linearly
with $\mu_B$ for $\mu_B \lesssim M_{\pi}$
while that of baryon decreases linearly~\cite{HSS}.
For $\mu_B \gtrsim M_{\pi}$, 
the baryon mass is not yet clearly confirmed.
On the other hand,
the behaviors of the masses
were also studied in an effective model~\cite{LSS}.
It is interesting to study the masses in a
general
effective model which can include several models
with the parameters chosen suitably.

In this paper 
we construct a low energy effective Lagrangian for the
two-color QCD
including the ``vector'' bosons
(mesons with $J^P=1^-$ and diquark baryons with $J^P=1^+$)
in addition to the pseudo-NG bosons (mesons with $J^P=0^-$ and
baryons with $J^P=0^+$).
The effective Lagrangian is composed 
based on the chiral symmetry breaking pattern of 
$SU(2N_f) \rightarrow Sp(2N_f)$
in the framework of the hidden local symmetry (HLS)~\cite{BKUYY,BKY,HY-PR},
and the ``vector'' bosons are introduced as the gauge bosons
of the $Sp(2N_f)$ HLS.
The HLS is equivalent to other 
models for ``vector'' bosons such as the CCWZ matter field~\cite{Matt},
the tensor field~\cite{Mass} and 
the Massive Yang-Mills field~\cite{Tens}.
Furthermore, it is possible to perform the systematic derivative
expansion in the HLS~\cite{Georgi,HY-PLB,Tanab,HY-PR}
, which allows us to study the parameter
dependences of the masses at finite density in a systematic way.

We study the vacuum structure of the model in the case of 
$N_f=2$ using the leading order Lagrangian, and
show that the flavor-singlet ``$\omega$''-meson carrying $J^P=1^-$
has a vacuum expectation value in the
time component. 
As a result,
the phase structure is the same as the one
determined by including only the pseudo-NG bosons~\cite{KSTVZ}:
For $\mu_B > M_{\pi}$, a baryonic pseudo-NG boson
($J^P=0^+$ state) condenses,
which causes the spontaneous breaking of the baryon number
symmetry, $U(1)_B$.
We show that
the mass of the anti-baryon (baryon)
with $J^P = 1^+$ increases (decreases) for $\mu_B < M_{\pi}$
and turns to decrease (increase) for $\mu_B > M_{\pi}$.
These behaviors
signal the phase transition of $U(1)_B$ breaking.
The effect of higher order terms
is shown to make $\rho$ and $\omega$ meson masses
decrease for $\mu_B > M_{\pi}$
consistently with lattice data~\cite{MNN,HSS}.
Furthermore,
the mass difference between $\rho$ and
$\omega$ mesons is proportional to the
mixing strength between the diquark baryon with $J^P=1^+$
and the anti-baryon.

The paper is organized as follows.
In section~\ref{sec:HLS}
we construct the chiral Lagrangian based on the HLS.
The vacuum structure and the $\mu_B$-dependences of the masses 
are studied at the leading order in section~\ref{sec:tree-mass}.
In section~\ref{sec:higher-mass} we show the effects of higher order terms
to the masses and mixings.
Section~\ref{sec:summary} is devoted to a summary and discussions.
Several intricate calculations and useful formulas are 
summarized in Appendices~\ref{sec:qcd-lag}-\ref{sec:O(p4)-lag}.

\section{HLS model in two-color QCD}
\label{sec:HLS}

Let us construct a low energy effective Lagrangian
including NG bosons associated with the spontaneous chiral symmetry
breaking $SU(2N_f) \rightarrow Sp(2N_f)$,
following
Ref.~\cite{KSTVZ}.

In the following we divide
the hermitian generators,
$\{ T^A \}$ of $SU(2N_f)$ normalized as 
$\tr [T^A T^B] = \delta^{AB}/2$,
into two classes:
The generators of $Sp(2N_f)$ denoted by $\{ S^{\alpha} \}$ with
$\alpha = 1,...,2N^2_f + N_f$;
and the remaining generators of $SU(2N_f)$ by
$\{ X^a \}$ with
$a = 1,...,2N^2_f - N_f - 1$.
These generators satisfy the relations
\begin{equation}
 (S^{\alpha})^T = \bar{\Sigma} S^{\alpha} \bar{\Sigma}, \qquad
 (X^a)^T = - \bar{\Sigma} X^a \bar{\Sigma},
\label{21}
\end{equation}
where $\bar{\Sigma}$ is a $2N_f \times 2N_f$ matrix
satisfying following properties:
\begin{equation}
 \bar{\Sigma}^2 = - \mbf{1}, \qquad
 \bar{\Sigma}^T = \bar{\Sigma}^{\dag} = - \bar{\Sigma}.
\label{22}
\end{equation}

The chiral symmetry breaking gives $2N^2_f - N_f - 1$
NG bosons $\pi$ which are encoded in the $2N_f \times 2N_f$ matrix as
\begin{equation}
 \Sigma = \xi(\pi) \bar{\Sigma} \xi^T(\pi),
\label{28}
\end{equation}
where
\begin{equation}
 \xi(\pi) = e^{i \pi / f_{\pi}}, \quad \pi = \pi^a X^a,
\label{29}
\end{equation}
with the decay constant $f_{\pi}$.~\footnote{
Note that the NG bosons consist of the mesons with $J^P = 0^{-}$ and the (anti-) baryons with $J^P = 0^{+}$.
}
Transformation property of $\xi(\pi)$ under the chiral symmetry is given by
\begin{equation}
 \xi (\pi) \rightarrow g \xi(\pi) h^{\dag}(\pi , g), \quad
 (h \in Sp(2N_f), \ g \in SU(2N_f)).
\label{213}
\end{equation}
From this together with the relations in Eq.~(\ref{21}),
we see that $\Sigma$ transforms linearly under the chiral symmetry as
\begin{equation}
 \Sigma \rightarrow g \Sigma g^T.
\label{214}
\end{equation}
The effective Lagrangian including the NG bosons
should be invariant under the global $SU(2N_f)$ group
and under the Lorentz transformation,
which is given by~\cite{KSTVZ}
\begin{equation}
 \mathcal{L} = \frac{f^2_{\pi}}{4} \tr
 [ (\partial_{\nu} \Sigma) ( \partial^{\nu} \Sigma^{\dag} ) ].
\label{215}
\end{equation}

Next, we include the ``vector'' boson fields~\footnote{
Here the ``vector'' boson 
fields imply the meson fields with $J^P = 1^-$
and the (anti-) baryon fields with $J^P = 1^+$.
} 
into the Lagrangian based on 
the hidden local symmetry (HLS)~\cite{BKUYY,BKY,HY-PR}.
We decompose the field $\Sigma$ as
\begin{equation}
 \Sigma
 = \xi^{\dag}_L \bar{\Sigma} \xi^T_L,
\label{217}
\end{equation}
where $\xi_L$ is given by
\begin{equation}
 \xi_L = \xi(\sigma) \xi^{\dag}(\pi), \quad
 (\, \xi(\sigma) = e^{i \sigma / f_{\sigma}},
  \quad \sigma = \sigma^{\alpha} S^{\alpha} \,),
\label{217a}
\end{equation}
with $\sigma$ being the NG bosons associated with the spontaneous
breaking of the HLS and $f_\sigma$ the corresponding decay constant.
The transformation property of $\xi_L$ is given by
\begin{equation}
 \xi_L \rightarrow h \xi_L g^{\dag},
\label{217b}
\end{equation}
where
\begin{equation}
 h \in [Sp(2N_f)]_{\textrm{local}}, \qquad
 g \in [SU(2N_f)]_{\textrm{global}}.
\label{219}
\end{equation}
For constructing the HLS Lagrangian,
it is convenient to introduce the field $\xi_R$ by
\begin{equation}
 \xi_R = \bar{\Sigma} (\xi^{\dag}_L(x))^T \bar{\Sigma}^{\dag}
 = \xi(\sigma) \xi(\pi),
\label{219a}
\end{equation}
which transforms as
\begin{equation}
 \xi_R \rightarrow h \xi_R ( \bar{\Sigma} g^T \bar{\Sigma}^{\dag} ).
\label{219b}
\end{equation}
Note that, in the HLS, the entire symmetry
$G_{\textrm{global}} \times H_{\textrm{local}}$
is spontaneously broken to its subgroup
$H_{\textrm{global}} = [Sp(2N_f)]_{\textrm{global}}$,
so that the NG bosons
$\sigma = \sigma^{\alpha} S^{\alpha}$
are absorbed into the HLS gauge bosons.
The basic quantities to construct the HLS Lagrangian
are the following two Maurer-Cartan 1-forms:
\begin{align}
 \hat{\alpha}_{\perp \nu}
 = [ ( D_{\nu} \xi_R ) \xi^{\dag}_R
   - ( D_{\nu} \xi_L ) \xi^{\dag}_L ]/(2i),
\label{223}
 \\
 \hat{\alpha}_{\parallel \nu}
 = [ ( D_{\nu} \xi_R ) \xi^{\dag}_R
   + ( D_{\nu} \xi_L ) \xi^{\dag}_L ]/(2i),
\label{224}
\end{align}
where the covariant derivatives are read
from the transformation properties in
Eqs.~(\ref{217b}) and (\ref{219b}) as
\begin{align}
 D_{\nu} \xi_L &= \partial_{\nu} \xi_L - i V_{\nu} \xi_L
 + i \xi_L G_{\nu},
\label{225}
 \\
 D_{\nu} \xi_R &= \partial_{\nu} \xi_R - i V_{\nu} \xi_R
 + i \xi_R ( \bar{\Sigma} G^T_{\nu} \bar{\Sigma} ),
\label{226}
\end{align}
with
$V_{\nu} = V^{\alpha}_{\nu} S^{\alpha}$
and
$G_{\nu} = G^{A}_{\nu} T^A$
being the $H_{\textrm{local}}$ gauge  bosons and
the external gauge bosons corresponding to the chiral symmetry
(see Appendix~\ref{sec:qcd-lag}),
respectively.
These gauge bosons transform as
\begin{align}
 V_{\nu} &\rightarrow V'_{\nu}
 = h V_{\nu} h^{\dag} + i h ( \partial_{\nu} h^{\dag} ),
\label{227}
 \\
 G_{\nu} &\rightarrow G'_{\nu}
 = g G_{\nu} g^\dag + i g (\partial_{\nu} g^{\dag}).
\label{228}
\end{align}
It should be noticed that in the HLS formalism
we can introduce the ``vector'' bosons $V_{\nu}$
and the external chiral gauge bosons $G_{\nu}$ independently.
Since the covariantized 1-forms 
$\hat{\alpha}_{\perp \nu}$ and
$\hat{\alpha}_{\parallel \nu}$
in Eqs.~(\ref{223}) and (\ref{224}) transform homogeneously:
\begin{equation}
 \hat{\alpha}^{\nu}_{\perp , \parallel} \rightarrow
h \hat{\alpha}_{\perp , \parallel}^{\nu}(x) h^{\dag},
\label{229}
\end{equation}
we have the following two invariants:
\begin{align}
 f^2_{\pi} \tr[ \hat{\alpha}_{\perp \nu} \hat{\alpha}_{\perp}^{\nu} ], \quad
 f^2_{\sigma}
 \tr[ \hat{\alpha}_{\parallel \nu} \hat{\alpha}_{\parallel}^{\nu} ].
\label{231}
\end{align}
We introduce the kinetic term of the ``vector'' bosons:
\begin{equation}
 - \frac{1}{2g^2} \tr [ V_{\nu \rho} V^{\nu \rho} ],
\label{234}
\end{equation}
where $g$ is the HLS gauge coupling constant
and $V_{\nu \rho}$ is the field strength defined by
\begin{equation}
 V_{\nu \rho} \equiv \partial_{\nu} V_{\rho} - \partial_{\rho} V_{\nu}
  -i [V_{\nu}, V_{\rho} ].
\label{232}
\end{equation}

In addition, we include
the external scalar and pseudoscalar source fields $\chi$
given in Eq.~(\ref{a7c}) into the Lagrangian.
Note that the vacuum expectation value (VEV) of $\chi$
gives the explicit chiral symmetry breaking
due to the current quark masses as
\begin{align}
 \langle \chi \rangle
 =
 \left(
  \begin{array}{ccc|ccc}
   {} & {} & {} & -m_{q_1} & {} & {} \\
   {} & \mbf{0}_{(N_f \times N_f)} & {} & {} & \ddots & {} \\
   {} & {} & {} & {} & {} & -m_{q_{N_f}} \\ \hline
   m_{q_1} & {} & {} & {} & {} & {} \\
   {} & \ddots & {} & {} & \mbf{0}_{(N_f \times N_f)} & {} \\
   {} & {} & m_{q_{N_f}} & {} & {} & {}
  \end{array}
 \right).
\label{235aa}
\end{align}
The $\chi$ transforms under the chiral symmetry as
$ \chi \rightarrow g^{\ast} \chi g^{\dag} $.
(see Eq.~(\ref{a7e}))
Since $\hat{\alpha}_{\parallel \nu}$ as well as $\hat{\alpha}_{\perp \nu}$
transforms homogeneously under the HLS,
it is convenient to convert $\chi$ into a field $\hat{\chi}$ as
\begin{align}
 \hat{\chi}
 =
 2 G \xi_R \bar{\Sigma} \chi \xi^{\dag}_L,
\label{235b}
\end{align}
where $G$ is a parameter carrying the mass dimension one.
$\hat{\chi}$ in Eq.~(\ref{235b}) transforms homogeneously
under the HLS
(see Eqs.~(\ref{217b}) and (\ref{219b})
for the transformation properties of $\xi_{L,R}$):
\begin{align}
 \hat{\chi}
 \rightarrow
 h \hat{\chi} h^{\dag}.
\label{235c}
\end{align}

For convenience, we summarize the transformation properties
of the building blocks under parity ($P$), charge-conjugation ($C$),
and HLS in TABLE~\ref{tab:trance}.
\begin{table}[htbp]
 \begin{tabular}{cccc}
 \hline
 Building block & $P$ & $C$ & HLS \\
 \hline
 $\hat{\alpha}^{\nu}_{\parallel}$ &
 $(\Omega \Sigma_c) \hat{\alpha}_{\parallel \nu} (\Omega \Sigma_c)$ &
 $- \hat{\alpha}^{\nu T}_{\parallel}$ &
 $h \hat{\alpha}^{\nu}_{\parallel} h^{\dag}$ \\
 $\hat{\alpha}^{\nu}_{\perp}$ &
 $- (\Omega \Sigma_c) \hat{\alpha}_{\perp \nu} (\Omega \Sigma_c)$ &
 $\hat{\alpha}^{\nu T}_{\perp}$ &
 $h \hat{\alpha}^{\nu}_{\perp} h^{\dag}$ \\
 $\hat{\chi}$ &
 $(\Omega \Sigma_c) \hat{\chi}^{\dag} (\Omega \Sigma_c)$ &
 $\hat{\chi}^T$ &
 $h \hat{\chi} h^{\dag}$ \\
 $G^{\nu}$ &
 $- \Omega G^T_{\nu} \Omega$ &
 $- \Sigma_c G^{\nu} \Sigma_c$ \, \, &
 $g G_{\nu} g^\dag + i g (\partial_{\nu} g^{\dag})$ \\
 \hline
 \end{tabular}
\caption[]{
Transformation properties
of the building blocks under parity ($P$), charge-conjugation ($C$),
and HLS.
}
\label{tab:trance}
\end{table}
In this table, $\Sigma_c$ is the $2N_f \times 2N_f$ matrix defined as
\begin{equation}
 \Sigma_c \equiv
 \left(
  \begin{array}{cc}
   \mbf{0} & \mbf{1}_N \\
   -\mbf{1}_N & \mbf{0}
  \end{array}
 \right),
\label{235e}
\end{equation}
which determines the direction of the chiral condensate, 
i.e. how to embed the $Sp(2N_f)$ into $SU(2N_f)$.
$\Omega$ is the $2N_f \times 2N_f$ matrix given as
\begin{equation}
 \Omega \equiv
 \left(
  \begin{array}{cc}
   \mbf{0} & \mbf{1}_N \\
   \mbf{1}_N & \mbf{0}
  \end{array}
 \right).
\label{235f}
\end{equation}
Then the lowest order term invariant under the $P$ transformation is
given by
\begin{align}
 \frac{f^2_{\chi}}{4} \tr [ \hat{\chi} + \hat{\chi}^{\dag} ].
\label{235d}
\end{align}
where $f_{\chi}$ is introduced
to renormalize the quadratically divergent correction
to this term~\cite{HY-PRD}.
In the present analysis,
we introduced this parameter
in such a way that the field $\hat{\chi}$ does not
get any renormalization effect.

Finally, the HLS Lagrangian with leading order terms is given by
\begin{align}
 \mathcal{L}
 = - \frac{1}{2g^2} \tr [ V_{\nu \rho} V^{\nu \rho} ]
    + f^2_{\sigma} \tr [ \hat{\alpha}^2_{\parallel \nu} ]
    + f^2_{\pi} \tr [ \hat{\alpha}^2_{\perp \nu}]
    + \frac{f^2_{\chi}}{4}
      \tr [ \hat{\chi} + \hat{\chi}^{\dag} ].
\label{239}
\end{align}

\section{Vacuum Structure and Spectrum of ``Vector'' bosons at nonzero baryon chemical potential}
\label{sec:tree-mass}

\begin{figure}
\begin{center}
\includegraphics[width=7cm,bb=147 140 428 327]{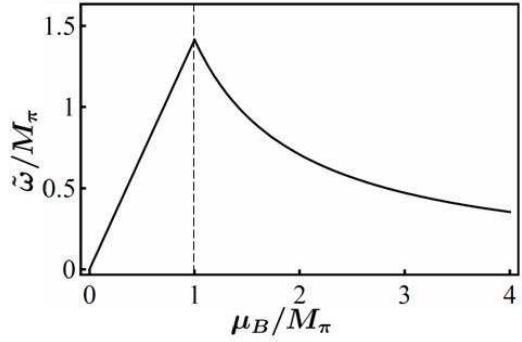}
\caption{$\mu_B$-dependence of the VEV $\tilde{\omega}$ in unit of $M_{\pi}$ as a function of $\mu_B / M_{\pi}$.}
\label{fig:omega-vev}
\end{center}
\end{figure}

In this section, we study the vacuum structure
and examine the effects of nonzero chemical potential
for the baryon number charge
on the spectrum of ``vector'' bosons in the case of $N_f = 2$.
In Table~\ref{tab:fields}, we show 
the fields included in the present model
together with their quantum numbers for the $SU(2)$
``isospin'' $I$~\footnote{
For $N_f = 2$ there exists an $SU(2)$ flavor symmetry
which we call the ``isospin'' symmetry.
}, 
the baryon number charge $B$~\footnote{
We follow the convention of baryon number
given in Ref.~\cite{KSTVZ},
which is different from the one in Ref.~\cite{HSS}.
}
and the 
spin-parity $J^P$.
\begin{table}[htbp]
 \begin{tabular}{ccccc}
 \hline
 Field  & Generator & $I$ \ \ \ & $B$ \ \ \ & $J^P$ \ \ \ \\
 \hline
 $\pi^{1,2,3}$ & $X^{1,2,3}$ & $1$ & $0$ & $0^-$ \\
 $\pi_{B_+} = (\pi^5 - i \pi^4)/\sqrt{2}$ & $(X^5 - i X^4)/\sqrt{2}$ &
 $0$ & $+1$ & $0^+$ \\
 $\pi_{B_-} = (\pi^5 + i \pi^4)/\sqrt{2}$ & $(X^5 + i X^4)/\sqrt{2}$ &
 $0$ & $-1$ & $0^+$ \\
 $\rho^{1,2,3} = V^{1,2,3}$ & $S^{1,2,3}$ & $1$ & $0$ & $1^-$ \\
 $\omega = V^4$ & $S^4$ & $0$ & $0$ & $1^-$ \\
 $V_{B_+} = (V^{\alpha} + i V^{\beta})/\sqrt{2}$ \ \ &
 $(S^{\alpha} + i S^{\beta})/\sqrt{2}$ \ & $1$ & $+1$ & $1^+$ \\
 $V_{B_-} = (V^{\alpha} - i V^{\beta})/\sqrt{2}$ \ \ &
 $(S^{\alpha} - i S^{\beta})/\sqrt{2}$ \ & $1$ & $-1$ & $1^+$ \\
 \hline
 \end{tabular}
\caption[]{
Fields corresponding to the mass eigenstates at $\mu_B=0$,
together with the ``isospin'' $I$, the baryon number charge $B$ and
the spin-parity $J^P$.
Indexes of the ``vector'' bosons with $J^P = 1^+$ are taken as
$(\alpha, \beta) = \{  (5,6), \, (7,8), \, (9,10)  \}$.
}\label{tab:fields}
\end{table}
The effect of chemical potential for the baryon number charge
is introduced as the vacuum expectation value (VEV)
of the external chiral gauge boson as
\begin{equation}
 \langle G_{\nu} \rangle
 = \delta_{0 \nu} \frac{\mu_B}{2}
 \left(
  \begin{array}{cc}
   \mbf{1} & \mbf{0} \\
   \mbf{0} & -\mbf{1}
  \end{array}
 \right)
 = \frac{\mu_B \delta_{0 \nu}}{2} B,
\label{238}
\end{equation}
where $\mbf{1}$ is the $2 \times 2$ unit matrix and
$
 \mbf{0}
 =
 \left(
  \begin{array}{cc}
   0 & 0 \\
   0 & 0
  \end{array}
 \right)
$.
In the following we restrict ourselves
to the case where two quarks have the same current quark masses:
The VEV of the external scalar field in Eq.~(\ref{235b}) takes
\begin{equation}
 \langle \chi \rangle
 = m_q
 \left(
  \begin{array}{cc}
   \mbf{0} & -\mbf{1} \\
   \mbf{1} & \mbf{0}
  \end{array}
 \right)
 = m_q \hat{M}.
\label{31a}
\end{equation}
Here we assume that the spatial rotational symmetry is not broken,
so that the relevant VEVs of the fields in the unitary gauge of the HLS,
$\sigma = 0$, are reduced to
\begin{equation}
 \langle \xi_{L} \rangle = \xi^{\dag}(\tilde{\pi}), \quad
 \langle \xi_{R} \rangle = \xi(\tilde{\pi}), \quad
 \langle V_{\nu}(x) \rangle
 = \tilde{V}_{\nu} = (\, \tilde{V}_0, \, \mbf{0} \,).
\label{31}
\end{equation}
Replacing the fields with three VEVs as above,
we obtain the static potential as
\begin{align}
 V_{\textrm{potential}}
 &=
 - f^2_{\sigma} \tr
 \left[
  \left\{
    \tilde{V}_0 - \frac{\mu_B}{2}
    \left( \,
     \xi^{\dag} (\tilde{\pi}) B \xi(\tilde{\pi}) +
     \xi(\tilde{\pi}) B \xi^{\dag} (\tilde{\pi})
    \, \right)
  \right\}^2
 \right]
 \nonumber\\
 &- \frac{f^2_{\pi} \mu^2_B}{16} \tr
 \left[
  \left\{
    \xi^{\dag} (\tilde{\pi}) B \xi(\tilde{\pi}) -
    \xi(\tilde{\pi}) B \xi^{\dag} (\tilde{\pi})
  \right\}^2
 \right]
 - \frac{M^2_{\pi} f^2_{\pi}}{4} \tr
 \left[
  \xi^2(\tilde{\pi}) \bar{\Sigma} \hat{M} +
  (\textrm{h.c})
 \right],
\label{32}
\end{align}
where we use $B^T = B$ and 
$M_{\pi}$ is the mass of $\pi$s defined as
\begin{align}
 M^2_{\pi} = \frac{2 G m_q f^2_{\chi}}{f^2_{\pi}}.
\label{33a}
\end{align}
From the stationary condition for $\tilde{V}_0$, we obtain
\begin{equation}
 \tilde{V}_0
 =
 \frac{\mu_B}{2}
 \left( \,
  \xi^{\dag}(\tilde{\pi}) B \xi(\tilde{\pi}) +
  \xi (\tilde{\pi}) B \xi^{\dag}(\tilde{\pi})
 \, \right).
\label{34}
\end{equation}
Substituting this into $V_{\textrm{potential}}$, we have
\begin{equation}
 V_{\textrm{potential}} \bigg|_{\tilde{V}_0}
 =
 -\frac{f^2_{\pi}} {4} \tr
 \left[
  \frac{\mu^2_B}{4}
  \left\{
    \xi^{\dag} (\tilde{\pi}) B \xi(\tilde{\pi}) -
    \xi(\tilde{\pi}) B \xi^{\dag} (\tilde{\pi})
  \right\}^2
  + M^2_{\pi} 
  \left\{
    \xi^2(\tilde{\pi}) \bar{\Sigma} \hat{M} +
   (\textrm{h.c})
  \right\}
 \right].
\label{35}
\end{equation}
This is equivalent to the potential term of the chiral Lagrangian
for NG bosons analyzed in
Ref.~\cite{KSTVZ}.
Then, the value of $\xi (\tilde{\pi})$
which minimizes the potential is obtained as
\begin{equation}
 \xi(\tilde{\pi})
 =
 e^{i \tilde{\pi}^5 X^5}
\label{310}
\end{equation}
where $X^5$ is given in Eqs.~(\ref{b6}) and (\ref{b7})
and the VEV $\tilde{\pi}^5$ is determined as
\begin{equation}
 \tilde{\pi}^5 = \sqrt{2} f_{\pi} \theta,
\label{311}
\end{equation}
with
\begin{equation}
 \theta
 =
 0, \quad (\textrm{for} \, 0 < \mu_B < M_{\pi}), \quad
 \cos \theta
 =
 \frac{M^2_{\pi}}{\mu^2_B}, \quad (\textrm{for} \, \mu_B > M_{\pi} ).
\label{39}
\end{equation}
This means
that there is a condensation of the baryonic-NG boson with 
$J^P = 0^{+}$
and $U(1)_B$ symmetry is broken spontaneously for $\mu_B > M_{\pi}$.
By substituting this VEV into Eq.~(\ref{34}), the VEV of the
``vector'' boson fields
$\tilde{V}^{\alpha}_0$ is determined as
\begin{align}
 \tilde{V}^{\alpha}_0 =
 \left\{
 \begin{array}{ll}
  \sqrt{2} \mu_B \cos \theta, \qquad &(\, \alpha = 4 \,), \\
  0, \qquad &(\, \alpha \neq 4 \,).
 \end{array}
 \right.
\label{312}
\end{align}
This implies that the time component of $\omega$ meson
(see Table~\ref{tab:fields})
has a VEV for any $\mu_B$, as in the ordinary three-color QCD.
In FIG.~\ref{fig:omega-vev}, we plot the $\mu_B$-dependence of
$\tilde{\omega} = \tilde{V}^4_0$.
This shows that the $\tilde{\omega}$ increases with $\mu_B$
for $\mu_B < M_{\pi}$ while it decreases for $\mu_B > M_{\pi}$.

We define the masses of ``vector'' bosons as the energies
determined from the
zero momentum limit of the dispersion relation.
We shift the ``vector'' boson fields as
\begin{equation}
 V_{\nu}(x) \rightarrow V_{\nu}(x) + \tilde{V}_{\nu},
\label{313b}
\end{equation}
and retain quadratic terms only in  Eq.~(\ref{239}).
Performing the Fourier transformation
and taking the low-momentum limit yields
\begin{align}
 &\int d^4 x \,
 \left[ \,
           - \frac{1}{2g^2} \tr ( V_{\nu \rho} V^{\nu \rho} )
           + f^2_{\sigma} \tr ( V_{\nu} V^{\nu} )
 \, \right]
 \nonumber\\
 &\rightarrow
 \frac{1}{2g^2} \hat{V}^{\alpha}_j (-E)
 \left[
  E^2 \delta^{\alpha \beta}
  - 2i f^{\alpha \beta \gamma} (\tilde{V}_0)^{\gamma} E
  + f^{\alpha \gamma \theta} f^{\beta \delta \gamma} (\tilde{V}_0)^{\gamma}
  (\tilde{V}_0)^{\delta}
  - M^2_V \delta^{\alpha \beta}
 \right] \hat{V}^{\beta}_j (E)
 \nonumber\\
 &\equiv \hat{V}^{\alpha}_j \Gamma^{\alpha \beta} \hat{V}^{\beta}_j,
\label{313}
\end{align}
where $f^{\alpha \beta \gamma}$ is a structure constant of $Sp(4)$,
$M_V$ is the mass of ``vector'' boson at $\mu_B = 0$;
\begin{equation}
 M_V \equiv g f_{\sigma},
\label{313d}
\end{equation}
and $\hat{V}$ is defined as
\begin{align}
 V^{\alpha}_{\nu}(x)
 =
 \int \frac{d^4 k}{(2 \pi)^4} \hat{V}^{\alpha}_{\nu} (k)
 e^{-i k x}.
\label{313a}
\end{align}

Using the basis of $Sp(4)$ generators shown in Appendix~\ref{sec:generators},
one finds that
the matrix $\Gamma$ for the inverse propagator at zero momentum limit
is block diagonal
with four 1$\times$1 terms and three 2$\times$2 blocks.
The four diagonal $1 \times 1$ terms are composed of
$\hat{V}^{1}$, $\hat{V}^{2}$, $\hat{V}^{3}$
corresponding to the $\rho$ meson and $\hat{V}^{4}$
corresponding to the $\omega$ meson.
The masses of these states are obtained as
\begin{equation}
  m_{\rho, \omega} = M_V \ .
\label{313e}
\end{equation}
Three 2 $\times$ 2 blocks for
$(\hat{V}^5, \hat{V}^6)$, $(\hat{V}^7, \hat{V}^8)$ and
$(\hat{V}^9, \hat{V}^{10})$
are identical because of the ``isospin'' symmetry.
The 2$\times$2 block for $\hat{V}^{5}$ and $\hat{V}^{6}$ is
obtained as
\begin{equation}
 (\hat{V}^{5 \dag}_j \hat{V}^{6 \dag}_j)
 \left(
  \begin{array}{cc}
   E^2 - M^2_V + \mu^2_B \cos^2 \theta & 2i \mu_B E \cos \theta \\
   -2i \mu_B E \cos \theta & E^2 - M^2_V + \mu^2_B \cos^2 \theta
  \end{array}
 \right)
 \left(
  \begin{array}{c}
   \hat{V}^5_j \\
   \hat{V}^6_j
  \end{array}
 \right).
\label{313f}
\end{equation}
This matrix is diagonalized by the
fields $V_{B_+}$ and $V_{B_-}$ defined in 
Table~\ref{tab:fields} as
\begin{equation}
 (\hat{V}^{\dag}_{B_{+},j} \hat{V}^{\dag}_{B_{-},j})
 \left(
  \begin{array}{cc}
   (E + \mu_B \cos \theta)^2 - M^2_V & 0 \\
   0 & (E - \mu_B \cos \theta)^2 - M^2_V
  \end{array}
 \right)
 \left(
  \begin{array}{c}
   \hat{V}_{B_{+},j} \\
   \hat{V}_{B_{-},j}
  \end{array}
 \right).
\label{313g}
\end{equation}
From this, we obtain the masses as
\begin{align}
 m_{V_{B+}} = M_V - \mu_B \cos \theta, \quad
 m_{V_{B-}} = M_V + \mu_B \cos \theta,
\label{319e}
\end{align}
where we assumed $M_V > M_{\pi}$,
so that $M_V > \mu_B \cos \theta$.
\begin{figure}
\begin{center}
\includegraphics[width=7cm,bb=147 157 428 349]{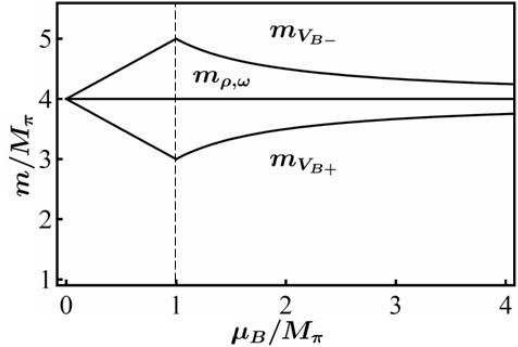}
\caption{The masses of the ``vector'' bosons
in unit of $M_{\pi}$ as a function of $\mu_B / M_{\pi}$. 
We use $M_V / M_{\pi}=4$ and $g=1$ to make the plot.} 
\label{fig:v-mass1}
\end{center}
\end{figure}

Let us consider the $\mu_B$-dependences of the masses of 
the ``vector'' bosons.
Equation~(\ref{313e}) indicates that the ordinary vector mesons with 
$J^P = 1^-$ do not change their masses at all.
On the other hand, from Eq.~(\ref{319e}) together with Eq.~(\ref{39}),
we find that the mass of the baryon ($V_{B_+}$)
with $J^P = 1^+$ decreases
for $\mu_B < M_{\pi}$
and turns to increase for $\mu_B > M_{\pi}$,
and the mass of anti-baryon ($V_{B_-}$) with $J^P = 1^+$
shows the opposite behavior.
This indicates that the phase transition can be observed by
seeing the masses of baryons with $J^P = 1^+$.
We stress that,
for any value of $\mu_B$,  the mass eigenstates are
given by $V_{B_+}$ and $V_{B_-}$, which are nothing but the
eigenstates of the baryon number;
$V_{B_+}$ carries $B=+1$ and $V_{B_-}$ does $B=-1$.
In other words,
the baryon with $J^P=1^+$ does not mix with
the anti-baryon having the same spin and parity, even though
the baryon number $U(1)_B$ symmetry is spontaneously broken
for $\mu_B > M_{\pi}$.
Note that this feature holds only at the leading order:
The mixing will generally appear
when we include the higher order terms
(see the next section).
In FIG.~\ref{fig:v-mass1},
we plot the masses of ``vector'' bosons
described in Eqs.~(\ref{313e}) and (\ref{319e})
for $M_V/M_{\pi} = 4$
as an example.
In this figure,
we can see the $\mu_B$-dependence of
the masses of the ``vector'' bosons explained above.

\section{Effect of higher order terms}
\label{sec:higher-mass}
In this section, we consider the effects of higher order terms
in the hidden local symmetry (HLS).
In QCD with three colors it is known that,
thanks to the gauge invariance of the HLS,
we can perform the systematic derivative expansion
with including vector mesons in addition to the pseudo Nambu-Goldstone bosons
when the masses of vector mesons are lighter
than the chiral symmetry breaking scale
(the chiral perturbation theory with the HLS~\cite{Georgi,HY-PLB,Tanab,HY-PR}).

We adopt the same counting rule in the present case.
Generally, there are 32 terms in the $\mathcal{O} (p^4)$
HLS Lagrangian~\cite{Tanab,HY-PR}.
We present a complete list of $\mathcal{O} (p^4)$ Lagrangian 
in Appendix~\ref{sec:O(p4)-lag}.
Here we include only the terms
which do not alter the vacuum structure given in Eq.~(\ref{35}),
neglecting the effect of current quark masses at $\mathcal{O} (p^4)$.
In this case we have only three combinations
which give corrections to the ``vector'' boson masses:
\begin{align}
 \mathcal{L}_{(4)_1}
 &=
 \bar{y}_1 \tr
 [ \,
  \hat{\alpha}_{\perp \mu} \hat{\alpha}_{\parallel \nu}
  \hat{\alpha}^{\mu}_{\perp} \hat{\alpha}^{\nu}_{\parallel}
  -
  \hat{\alpha}_{\perp \mu} \hat{\alpha}^{\mu}_{\perp}
  \hat{\alpha}_{\parallel \nu} \hat{\alpha}^{\nu}_{\parallel}
  +
  \hat{\alpha}_{\perp \mu} \hat{\alpha}_{\perp \nu}
  \hat{\alpha}^{\mu}_{\parallel} \hat{\alpha}^{\nu}_{\parallel}
 \nonumber\\
  &\qquad
  -\frac{1}{2}
  (
   \hat{\alpha}_{\perp \mu} \hat{\alpha}^{\mu}_{\parallel}
   \hat{\alpha}_{\perp \nu} \hat{\alpha}^{\nu}_{\parallel}
   +
   \hat{\alpha}_{\perp \mu} \hat{\alpha}_{\parallel \nu}
   \hat{\alpha}^{\nu}_{\perp} \hat{\alpha}^{\mu}_{\parallel}
  )
 \, ],
\label{41a}\\
 \mathcal{L}_{(4)_2}
 &=
 \bar{y}_2 \tr
 [ \,
  \hat{\alpha}_{\perp \mu} \hat{\alpha}_{\parallel \nu}
  \hat{\alpha}^{\mu}_{\perp} \hat{\alpha}^{\nu}_{\parallel}
  -
  \hat{\alpha}_{\perp \mu} \hat{\alpha}_{\perp \nu}
  \hat{\alpha}^{\nu}_{\parallel} \hat{\alpha}^{\mu}_{\parallel}
  +
  \hat{\alpha}_{\perp \mu} \hat{\alpha}_{\perp \nu}
  \hat{\alpha}^{\mu}_{\parallel} \hat{\alpha}^{\nu}_{\parallel}
 \nonumber\\
  &\qquad
  -\frac{1}{2}
  (
   \hat{\alpha}_{\perp \mu} \hat{\alpha}^{\mu}_{\parallel}
   \hat{\alpha}_{\perp \nu} \hat{\alpha}^{\nu}_{\parallel}
   +
   \hat{\alpha}_{\perp \mu} \hat{\alpha}_{\parallel \nu}
   \hat{\alpha}^{\nu}_{\perp} \hat{\alpha}^{\mu}_{\parallel}
  )
 \, ],
\label{41b}\\
 \mathcal{L}_{(4)_3}
 &=
 \bar{y}_3 \tr
 [ \,
  \hat{\alpha}_{\perp \mu} \hat{\alpha}_{\parallel \nu}
  \hat{\alpha}^{\mu}_{\perp} \hat{\alpha}^{\nu}_{\parallel}
  -
  \hat{\alpha}_{\perp \mu} \hat{\alpha}^{\mu}_{\perp}
  \hat{\alpha}_{\parallel \nu} \hat{\alpha}^{\nu}_{\parallel}
  +
  \hat{\alpha}_{\perp \mu} \hat{\alpha}_{\perp \nu}
  \hat{\alpha}^{\nu}_{\parallel} \hat{\alpha}^{\mu}_{\parallel}
 \nonumber\\
  &\qquad
  -\frac{1}{2}
  (
   \hat{\alpha}_{\perp \mu} \hat{\alpha}^{\mu}_{\parallel}
   \hat{\alpha}_{\perp \nu} \hat{\alpha}^{\nu}_{\parallel}
   +
   \hat{\alpha}_{\perp \mu} \hat{\alpha}_{\parallel \nu}
   \hat{\alpha}^{\nu}_{\perp} \hat{\alpha}^{\mu}_{\parallel}
  )
 \, ],
\label{41c}
\end{align}
where $\bar{y}_1$, $\bar{y}_2$ and $\bar{y}_3$
are coefficients not determined by the HLS.

We consider the case of $N_f = 2$ in the following analysis.
From the vacuum expectation values (VEVs)
of $\xi(\pi)$ and $V_{\nu}$
given in Eqs.~(\ref{310}) and (\ref{312}),
VEVs of $\hat{\alpha}_{\parallel \mu}$ and $\hat{\alpha}_{\perp \mu}$
are determined as
\begin{align}
 \langle \hat{\alpha}_{\parallel \mu} \rangle
 =
 0, \quad
 \langle \hat{\alpha}_{\perp \mu} \rangle
 =
 \sqrt{2} \mu_B \delta_{\mu 0} \sin \theta X^4,
\label{42}
\end{align}
where $X^4$ is the fourth component of the broken generators
given in Eq.~(\ref{b6}).
Substituting the VEV $\langle \hat{\alpha}_{\perp \mu} \rangle$
into Eqs.~(\ref{41a})-(\ref{41c})
we obtain the correction to the masses of ``vector'' bosons as
\begin{align}
 \mathcal{L}_{(4)_1} + \mathcal{L}_{(4)_2} + \mathcal{L}_{(4)_3}
 \stackrel{\mathcal{O} (V^2_j)}{\rightarrow}
 2 \mu^2_B \sin^2 \theta
 [
  C_1 \tr (X^4 V_j X^4 V^j) - C_2 \tr (X^4 X^4 V_j V^j)
 ],
\label{43}
\end{align}
where $C_1$ and $C_2$ are certain linear combinations of
$\bar{y}_1$, $\bar{y}_2$ and $\bar{y}_3$.
\begin{figure}
\begin{center}
\includegraphics[width=14cm,bb=51 162 522 322]{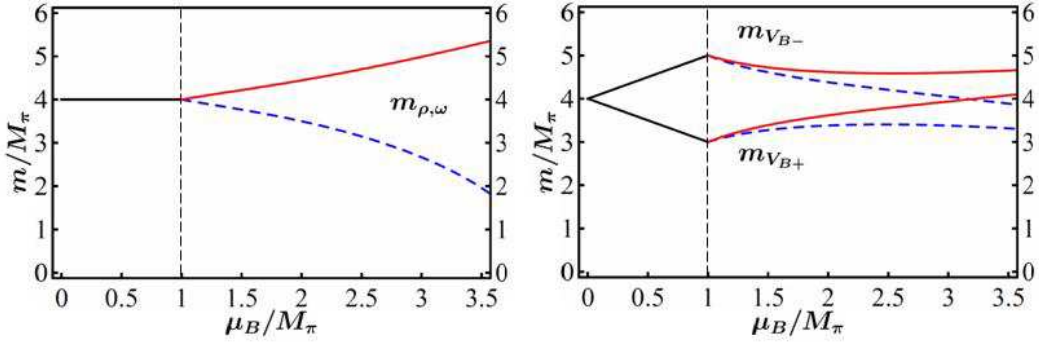}
\caption{
$\mu_B$-dependences of the ``vector'' boson masses.
The curves on the left figure are for the degenerate $\rho$ and $\omega$
mesons and those on the right figure are for baryonic and 
anti-baryonic ``vector'' bosons.
We use $M_V / M_{\pi} = 4$, $C_1 = 0$, $C_2 = \pm 4$ and $g=1$
to make the curves.
Dashed (blue) curves stand for $C_2 = 4$.
Solid (red) curves stand for $C_2 = -4$.
}
\label{fig:v-mass2}
\end{center}
\end{figure}
Applying the same procedure as that in section~\ref{sec:tree-mass}
with the correction in Eq.~(\ref{43}),
we obtain the masses of $\rho$ meson and $\omega$ meson as
\begin{align}
 &m_{\rho}
 =
 \sqrt{M^2_V + \frac{C_1 - C_2}{4} (g \mu_B \sin \theta)^2},
\label{45}\\
 &m_{\omega}
 =
 \sqrt{M^2_V - \frac{C_1 + C_2}{4} (g \mu_B \sin \theta)^2}.
\label{46}
\end{align}
The quadratic term for $(V_{B_+} , V_{B_-})$ is obtained as
{\footnotesize
\begin{align}
 (\hat{V}^{\dag}_{B_{+}} \hat{V}^{\dag}_{B_{-}})
 \left(
  \begin{array}{cc}
   (E + \mu_B \cos \theta)^2 - M^2_V + \frac{C_2}{16} (g \mu_B \sin \theta)^2 &
   \frac{C_1}{16} (g \mu_B \sin \theta)^2 \\
   \frac{C_1}{16} (g \mu_B \sin \theta)^2 &
   (E - \mu_B \cos \theta)^2 - M^2_V + \frac{C_2}{16} (g \mu_B \sin \theta)^2
  \end{array}
 \right)
 \left(
  \begin{array}{c}
   \hat{V}_{B_{+}} \\
   \hat{V}_{B_{-}}
  \end{array}
 \right).
\label{46a}
\end{align}
}%
The masses of ``vector'' bosons with $J^P = 1^+$
are obtained by diagonalizing the mass matrix
in Eq.~(\ref{46a}).
We can see that
the mixing between $V_{B_+}$ and $V_{B_-}$ is related with
the mass difference between $m_{\rho}$ and $m_{\omega}$ as
\begin{align}
 m^2_{\rho} - m^2_{\omega}
 =
 \frac{C_1}{2} (g \mu_B \sin \theta)^2.
\label{47}
\end{align}

Equations~(\ref{45}), (\ref{46}) and (\ref{46a}) show that
the corrections to the masses from ${\mathcal O}(p^4)$ terms
include the factor of $\sin\theta$, which is zero for
$\mu_B < M_{\pi} $.
Then, the ${\mathcal O}(p^4)$ corrections appear 
only for $\mu_B > M_{\pi}$,
where the $U(1)_B$ is spontaneously broken.
It should be noticed that two coefficients $C_1$ and $C_2$
are not determined by the symmetry structure.
To study the effect of ${\mathcal O}(p^4)$ corrections
we refer to the mass spectra obtained 
by the lattice simulation in Ref.~\cite{HSS},
which shows that 
the masses of $\rho$ and $\omega$ mesons are degenerate,
and that both of them
are stable against the change of the chemical potential
for $\mu_B < M_{\pi}$
and decrease for $\mu_B > M_{\pi}$.
From Eqs.~(\ref{45}) and (\ref{46}) the degeneracy 
of $m_{\rho}$ and $m_{\omega}$
is realized for $C_1=0$,
and both $m_{\rho}$ and $m_{\omega}$ decrease for $C_2 > 0$.
From Eq.~(\ref{43}) together with $X^4 X^4 = 1$,
the choice of $C_1 = 0$ and $C_2 > 0$
implies that the ${\mathcal O}(p^4)$ terms
provide a negative contribution to all the ``vector'' boson masses
equally, and that
$V_{B_+}$ and $V_{B_-}$ do not mix
with each other, even though the baryon number
$U(1)_B$ symmetry is spontaneously broken.
On the other hand,
the choice of $C_1 = 0$ and $C_2 < 0$
implies that the ${\mathcal O}(p^4)$ terms
provide a positive contribution to all the ``vector'' boson masses
equally
(see FIG.~\ref{fig:v-mass2}).
The effect of nonzero $C_1$ produces the mass difference
between $m_{\rho}$ and $m_{\omega}$,
and this difference is linked to the mixing strength
between $V_{B_+}$ and $V_{B_-}$ as in Eq.~(\ref{47}).
This relation is obtained from
the symmetry breaking pattern
and the assumption that
all the bosons other than $\pi$ and $V$ are heavy enough
to be neglected in the Lagrangian.~\footnote{
We expect that
the contribution of higher order terms such as $\mathcal{O} (p^6)$
is small enough.
}
Thus, a violation of Eq.~(\ref{47}),
when only $\pi$ and $V$ are light degrees of freedom,
may signal a new phase transition.

\begin{figure}
\begin{center}
\includegraphics[width=14cm,bb=44 162 508 314]{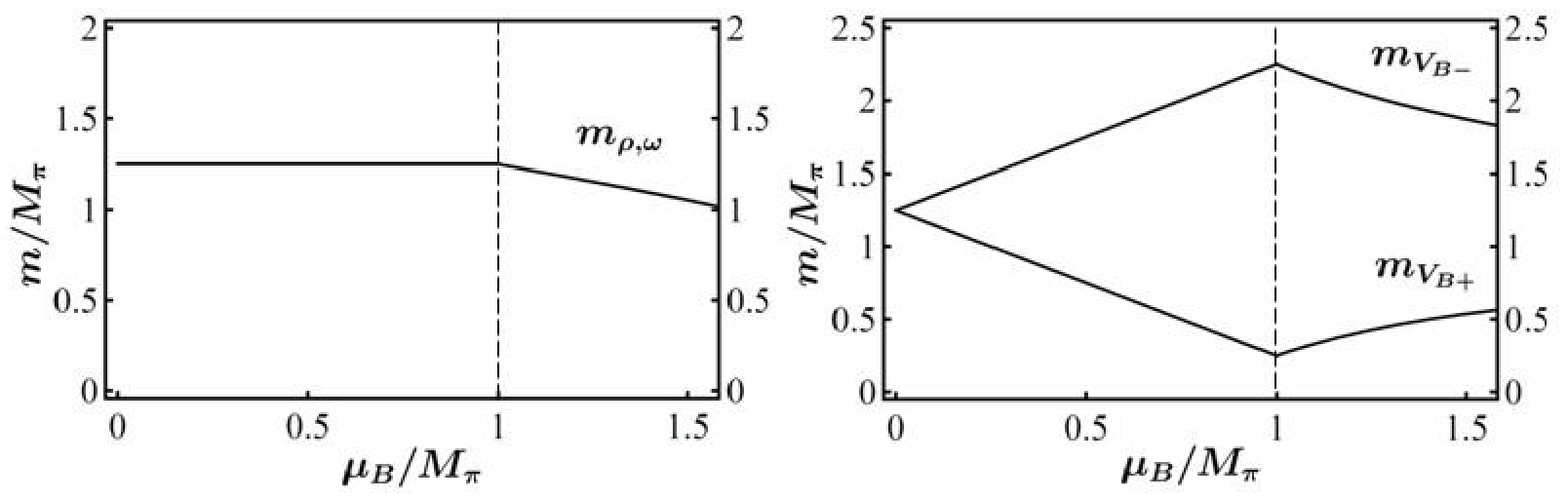}
\caption{
$\mu_B$-dependences of the ``vector'' boson masses.
The curve on the left figure is for the degenerate $\rho$ and $\omega$
mesons and those on the right figure are for baryonic and 
anti-baryonic ``vector'' bosons. We use $M_V / M_{\pi} = 5 / 4$, $C_1 = 0$, $C_2 = 1$ and $g=1$ to make the curves.
}
\label{v-mass3}
\end{center}
\end{figure}
We consider the case with $M_V/M_{\pi} = 5/4$, i.e., relatively
heavy $\pi$s,
which corresponds to the one in the lattice analysis~\cite{HSS}.
As an example, 
we plot the $\mu_B$-dependence of the ``vector'' boson masses
for $C_1 = 0$ and $C_2 = 1$ together with $M_V/M_{\pi} = 5/4$ and $g=1$
in FIG.~\ref{v-mass3}.
We choose the value $C_2 = 1$ to reproduce
the decreasing masses of $\rho$ and $\omega$ mesons
in the lattice data~\cite{HSS} for $\mu_B > M_{\pi}$.
We obtain the result that
$m_{V_{B+}}$ increases
and $m_{V_{B-}}$ decreases with increasing $\mu_B$
for $\mu_B > M_{\pi}$.
The lattice analysis~\cite{HSS} shows that
$m_{V_{B+}}$ is almost stable against the change of $\mu_B$,
which does not agree with the result of the present analysis.
Though any clear signal for $m_{V_{B-}}$ is not observed.

\begin{figure}
\begin{center}
\includegraphics[width=14cm,bb=56 165 521 324]{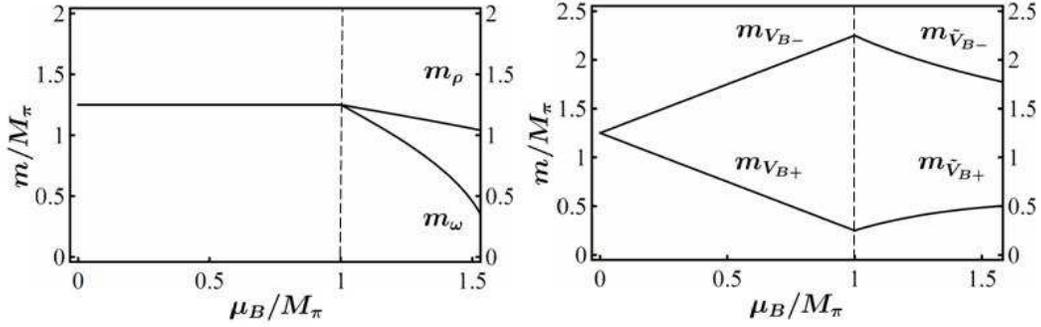}
\caption{
$\mu_B$-dependences of the ``vector'' boson masses.
The curves on the left figure are for the $\rho$ and $\omega$
mesons and those on the right figure are for baryonic and 
anti-baryonic ``vector'' bosons.
We use $M_V / M_{\pi} = 5/4 $, $C_1 = 1$, $C_2 = 2$ and
$g=1$ to make the curves.
}
\label{fig:v-mass4}
\end{center}
\end{figure}

We investigate the effect of $C_1$ term
to $m_{V_{B+}}$ and $m_{V_{B-}}$ in Eq.~(\ref{46a}).
This term causes a mixing between $V_{B_+}$ and $V_{B_-}$,
and makes $m_{V_{B+}}$ ($m_{V_{B-}}$) smaller (larger).
At the same time,
the $C_1$ term produces the mass difference
between the $\rho$ and $\omega$ mesons (see Eq.~(\ref{47})):
The positive $C_1$ gives the positive correction to 
$m_{\rho}$ and the negative one to $m_{\omega}$,
and the negative $C_1$ gives the negative correction to 
$m_{\rho}$ and the positive one to $m_{\omega}$.
As an example, we plot the 
$\mu_B$-dependence of the ``vector'' boson masses for 
$C_1 = 1$ and $C_2 = 2$ together with $M_V/M_{\pi} = 5/4$ and $g=1$
in FIG.~\ref{fig:v-mass4}.
We set $C_2 - C_1 = 1$ to reproduce
the decreasing mass of $\rho$ meson
in the lattice data~\cite{HSS},
and $C_1 = 1$ to produce 10\% decreasing of $m_{V_{B+}}$
at $\mu_B / M_{\pi} = 1.5$.
Left panel of FIG.~\ref{fig:v-mass4} shows that
the splitting between $m_{\rho}$ and $m_{\omega}$ is large:
$m_{\omega}$ is about half of $m_{\rho}$ at $\mu_B / M_{\pi} = 1.5$.
Thus, our model cannot simultaneously reproduce the $\mu_B$-dependences
of $m_{\rho}$, $m_{\omega}$ and $m_{V_{B+}}$ in the lattice data~\cite{HSS}.

For relatively heavy $\pi$s, we expect that the masses of
``axial-vector'' bosons are smaller than $M_V + M_{\pi}$ which is
the largest value of ``vector'' boson masses in the present analysis,
so that we cannot neglect the ``axial-vector'' bosons.
When we include the ``axial-vector'' meson $A$,
as well as the baryonic and anti-baryonic ``axial-vector'' bosons,
$A_{B_+}$ and $A_{B_-}$ (see Table~\ref{tab:axial}),
the $A$ will mix with baryonic and anti-baryonic
``vector'' bosons, $V_{B_+}$ and $V_{B_-}$,
and $A_{B_+}$ and $A_{B_-}$ will mix with $\omega$ meson
for $\mu_B > M_{\pi}$.
The effect of mixing among $A$, $V_{B_+}$ and $V_{B_-}$
will make $m_{V_{B+}}$ lighter
since $m_{V_{B+}}$ is the smallest among three at $\mu_B = M_{\pi}$.
As a result, the $m_{V_{B+}}$ becomes more stable than
that shown in FIG.~\ref{v-mass3},
and the lattice data will be reproduced.
This strongly suggests that
there is a large mixing among $A$, $V_{B_+}$ and $V_{B_-}$
for heavy $\pi$s such as the one adopted
in the lattice analysis~\cite{HSS}.
This indicates that
the mass of isovector meson with $J^P = 1^+$
shown in FIG.~2 of Ref.~\cite{HSS}
is nothing but $m_{V_{B+}}$ observed through the large mixing.
On the other hand,
the mixing among $A_{B_+}$, $A_{B_-}$ and $\omega$
will generally break the 
degeneracy between the $\rho$ and $\omega$ mesons
as shown in e.g. Ref.~\cite{LSS},
which does not seem to agree with the lattice result.
Then, the mixing expected to small enough to
reproduce the lattice result.

\begin{table}[htbp]
 \begin{tabular}{ccccc}
 \hline
 Field & Generator & $I$ \ \ \ \ \ \ & $B$ \ \ \ \ \ \ &
 $J^P$ \ \ \ \ \ \ \\
 \hline
 $A^{1,2,3}$ & $X^{1,2,3}$ & $1$ & $0$ & $1^+$ \\
 $A_{B_+} = (A^5 - i A^4)/ \sqrt{2}$ \ \ & $(X^5 - i X^4)/ \sqrt{2}$ \ &
 $0$ & $+1$ & $1^-$ \\
 $A_{B_-} = (A^5 + i A^4)/ \sqrt{2}$ \ \ & $(X^5 + i X^4)/ \sqrt{2}$ \ &
 $0$ & $-1$ & $1^-$ \\
 \hline
 \end{tabular}
\caption[]{
Summary of the ``axial-vector'' fields
}\label{tab:axial}
\end{table}

\section{Summary and discussions}
\label{sec:summary}

In this paper,
we constructed a low energy effective Lagrangian for the
two-color QCD
including the ``vector'' bosons
(mesons with $J^P=1^-$ and diquark baryons with $J^P=1^+$)
in addition to the pseudo Nambu-Goldstone bosons
(mesons with $J^P=0^-$ and baryons with $J^P=0^+$)
based on the chiral symmetry breaking pattern of 
$SU(2N_f) \rightarrow Sp(2N_f)$
in the framework of the hidden local symmetry (HLS).
The ``vector'' bosons were introduced as the gauge bosons
of the $Sp(2N_f)$ HLS.
In the HLS formalism,
the ``vector'' bosons and the external chiral gauge bosons
are included independently,
so that we can naturally incorporate the chemical potential $\mu_B$
as the vacuum expectation value (VEV)
of the external gauge boson for baryon number.

We studied the vacuum structure of the model in the case of $N_f=2$
introducing the effects of current quark masses and 
the baryon chemical potential into the leading order Lagrangian.
We found that
the time component of ``$\omega$'' meson
has a VEV for any $\mu_B$, as in the ordinary three-color QCD.
As a result,
the phase structure is the same as the one
determined by including only the pseudo-NG bosons~\cite{KSTVZ}:
For $\mu_B > M_{\pi}$, the baryonic pion ($J^P=0^+$ state) condenses,
which causes the spontaneous breaking of the baryon number
symmetry, $U(1)_B$.

We investigated the $\mu_B$-dependences of the
``vector'' boson masses
and found that
the mass of anti-baryon with $J^P = 1^+$ ($V_{B_-}$) increases for
$\mu_B < M_{\pi}$ and turns to decrease for $\mu_B > M_{\pi}$.
The mass of baryon with $J^P = 1^+$ ($V_{B_+}$)
shows the opposite behavior.
These behaviors of the baryons
signal the phase transition of $U(1)_B$ breaking
in two-color QCD at finite density.

At the leading order
the vector mesons with $J^P = 1^-$ ($\rho$ and $\omega$ mesons)
do not change their masses at all.
Furthermore,
$V_{B_-}$ does not mix with $V_{B_+}$,
even though the baryon number $U(1)_B$ symmetry is spontaneously broken
for $\mu_B > M_{\pi}$.

We studied the effects of higher order terms and
obtained the corrections to the masses of the ``vector'' bosons.
We assumed that the vacuum structure is not changed,
which left only two free parameters $C_1$ and $C_2$:
The positive (negative) $C_2$ provides
a negative (positive) contribution
to all the ``vector'' boson masses equally.
On the other hand,
the effect of nonzero $C_1$ produces the mass difference
between $m_{\rho}$ and $m_{\omega}$,
and this difference is linked to the mixing strength
between $V_{B_+}$ and $V_{B_-}$.
This relation is obtained from
the symmetry breaking pattern
and the assumption that
all the bosons other than $\pi$ and $V$ are heavy enough
to be neglected in the Lagrangian.
Thus, a violation of this relation (Eq.~(\ref{47}))
may signal a new phase transition.

Comparison with the lattice data in Ref.~\cite{HSS}
strongly suggests that
there is a large mixing among $A$, $V_{B_+}$ and $V_{B_-}$
for heavy $\pi$s such as the one adopted
in Ref.~\cite{HSS}.
This indicates that
the mass of isovector meson with $J^P = 1^+$
shown in FIG.~2 of Ref.~\cite{HSS}
is nothing but $m_{V_{B+}}$ observed through the large mixing.

We make a comment on the analysis done in 
Ref.~\cite{LSS}.
When the ``axial-vector'' bosons are taken to be
heavy in their model and integrated out, 
the model becomes identical to the HLS model with 
the parameter choice $C_1 = C_2 = 2 (M_S / M_X)^4 / g^2$,
where $M_S$ and $M_X$ express
the masses of ``vector'' bosons and ``axial-vector'' bosons 
in their model.
As can be seen easily from Eq.~(4.6), the choice $C_1 = C_2$ 
implies that the $\rho$ meson mass $m_\rho$ is stable against 
the density as shown in Ref.~\cite{LSS}.

The present analysis is valid when the
``axial-vector'' bosons
(mesons with $J^P=1^+$ and baryons with $J^P=1^-$)
are heavy.
When the ``axial-vector'' bosons are light,
we need to include these states.
This can be done in the framework of the
generalized HLS~\cite{BKY,BFY}.
Using this formalism,
we may investigate the phase structure
in the range of $\mu_B$ wider than that studied
in the present analysis.
We hope to obtain a clue to understand
the real-life QCD with three colors at finite baryon density
through these analyses.

\subsection*{Acknowledgments}

This work is supported in part by
Global COE Program ``Quest for
Fundamental Principles in the Universe'' of Nagoya
University (G07).
The work of M.H. is supported
in part by the JSPS Grant-in-Aid for Scientific Research
(c) 20540262 and
Grant-in-Aid for Scientific Research on Innovative
Areas (No. 2104)
``Quest on New Hadrons with Variety of Flavors''
from MEXT.
The work of C.N. is supported in part by the
Mitsubishi Foundation.

\appendix

\setcounter{section}{0}
\renewcommand{\thesection}{\Alph{section}}
\setcounter{equation}{0}
\renewcommand{\theequation}{\Alph{section}.\arabic{equation}}

\section{QCD Lagrangian with external source fields}
\label{sec:qcd-lag}
In this appendix, we give the
QCD Lagrangian with the external source fields.

We start with the ordinary QCD Lagrangian with $N_f$ massless quarks:
\begin{align}
 \mathcal{L}^0_{\textrm{QCD}}
 =
 - \frac{1}{2} \tr[ G^{\nu \rho} G_{\nu \rho} ]
 + \bar{\psi} \gamma^{\nu} D_{\nu} \psi,
\label{a2}
\end{align}
where
\begin{align}
 D_{\nu} \psi &= (\partial_{\nu} - i g_s G_\nu ) \psi,
 \nonumber\\
 G_{\nu \rho} &= \partial_{\nu} G_{\rho} - \partial_{\rho} G_{\nu}
 - i g_s [G_{\nu} \,  , \,  G_{\rho}],
\label{a2a}
\end{align}
with $G_{\nu}$ and $g_s$ being the gluon field matrix
and the gauge coupling constant.
Note that the gluon field matrix is expressed as
$G_\nu = G^a_{\nu} \frac{\tau_a}{2}$
where $\tau_a$ is the Pauli matrix of $SU(2)_{\textrm{color}}$ 
defined as
\begin{align}
 \tau_1 =
  \left(
   \begin{array}{cc}
    0 & 1  \\
    1 & 0
   \end{array}
  \right), \,
 \tau_2 =
  \left(
   \begin{array}{cc}
    0 & -i  \\
    i & 0
   \end{array}
  \right), \,
 \tau_3 =
  \left(
   \begin{array}{cc}
    1 & 0  \\
    0 & -1
   \end{array}
  \right).
\label{a3}
\end{align}

We include external scalar and pseudoscalar source fields
$\mathcal{S}$ and $\mathcal{P}$,
as well as
scalar and pseudoscalar diquark source fields
$\mathcal{Q}$ and $\mathcal{R}$~\footnote{
$\mathcal{Q}$ and $\mathcal{R}$ are color-singlet states
in two-color QCD.
This can be seen from the
existence of $i \tau_2 = \epsilon$ and
the charge-conjugation matrix $C \equiv i \gamma_2 \gamma_0$ 
in Eq.~(\ref{a3a}).
} as
\begin{align}
 \mathcal{L}_{\textrm{ext-scalar}}
 =
 \bar{\psi} (\mathcal{S} + i \mathcal{P}) \psi +
 \frac{1}{2}
 [
  \psi^T C (\gamma_5 \mathcal{Q} + i \mathcal{R}) i \tau_2 \psi +
  (\textrm{h.c})
 ].
\label{a3a}
\end{align}
The vector and axial-vector external gauge fields
$\mathcal{V}^{\mu}$ and $\mathcal{A}^{\mu}$
as well as
diquark external gauge fields carrying $J^P = 1^+$ and $1^-$
($\mathcal{B}^{\mu}$ and $\mathcal{D}^{\mu}$)
are included as
\begin{align}
 \mathcal{L}_{\textrm{ext-vector}}
 =	
 \bar{\psi} \gamma_{\mu}
 (\mathcal{V}^{\mu} - \gamma_5 \mathcal{A}^{\mu}) \psi +
 \frac{1}{2}
 [
  \psi^T C \gamma_{\mu} (\mathcal{B}^{\mu} - \gamma_5 \mathcal{D}^{\mu})
  i \tau_2 \psi +
  (\textrm{h.c})
 ].
\label{a3b}
\end{align}
These external fields satisfy the following conditions:
\begin{align}
 &\mathcal{S}^\dag = \mathcal{S}, \quad
 \mathcal{P}^\dag = \mathcal{P}, \quad
 (\mathcal{V}^{\mu})^{\dag} = \mathcal{V}^{\mu}, \quad
 (\mathcal{A}^{\mu})^{\dag} = \mathcal{A}^{\mu},
 \nonumber\\
 &\mathcal{Q}^T = - \mathcal{Q}, \quad
 \mathcal{R}^T = - \mathcal{R}, \quad
 (\mathcal{B}^{\mu})^T = \mathcal{B}^{\mu}, \quad
 (\mathcal{D}^{\mu})^T = - \mathcal{D}^{\mu}.
\label{a3c}
\end{align}
Now, the total Lagrangian is given by
\begin{equation}
 \mathcal{L}_{\textrm{QCD}}
 =
 \mathcal{L}^0_{\textrm{QCD}} +
 \mathcal{L}_{\textrm{ext-scalar}} +
 \mathcal{L}_{\textrm{ext-vector}} \, .
\label{a1}
\end{equation}

It is convenient to introduce
two-component spinors $q_L$ and $q_R$
in such a way that 
the four-component spinor $\psi$ is expressed as
$
 \psi^i =
  \left(
   \begin{array}{c}
    q^i_L  \\
    q^i_R
   \end{array}
  \right),
$
where $i$ denotes the flavor index.
Then, the kinetic term of quarks is rewritten as
\begin{equation}
 \int d^4 x \, i \bar{\psi}_i \gamma_{\nu} D^{\nu} \psi^i
 =
 \int d^4 x \,
 \bigl(
       i q^{\dag}_{L,i} \sigma_{\nu} D^{\nu} q^i_L
       + i q^{\dag}_{R,i} \bar{\sigma}_{\nu} D^{\nu} q^i_R
 \bigl),
\label{a4}
\end{equation}
where we use the following form of the gamma matrices:
\begin{equation}
 \gamma_{\nu}
 =
 \left(
  \begin{array}{cc}
   \mbf{0}      & \bar{\sigma}_{\nu} \\
   \sigma_{\nu} & \mbf{0}
  \end{array}
 \right)
 \equiv 
 \left(
  \begin{array}{cc}
   \mbf{0}               & (\mbf{1}, \sigma_j) \\
   (\mbf{1}, - \sigma_j) & \mbf{0}
  \end{array}
 \right).
\label{a5}
\end{equation}
{}From the form given in Eq.(\ref{a4}) 
the existence of the $SU(2N_f)$ flavor symmetry
is seen as follows:
Since the fundamental representation of $SU(2)_{\textrm{color}}$
as well as that of $SU(2)_{\textrm{spin}}$ is the 
pseudreal representation,
a combination of $\sigma_2 \tau_2 q^{\ast}_R$ has
the same transformation property as $q_L$
under the $SU(2)_{\textrm{color}}$ symmetry
as well as under the Lorentz symmetry.
Then, by introducing the field $\Psi$ as
\begin{equation}
 \Psi =
 \left(
  \begin{array}{c}
   q^1_L     \\
   q^2_L     \\
   \vdots    \\
   q^{N_f}_L \\
   \sigma_2 \tau_2 q^{\ast}_{R,1} \\
   \sigma_2 \tau_2 q^{\ast}_{R,2} \\
   \vdots                        \\
   \sigma_2 \tau_2 q^{\ast}_{R,N_f}
  \end{array}
 \right),
\label{a6}
\end{equation}
the kinetic term in Eq.(\ref{a4}) is rewritten as
\begin{equation}
 \int d^4 x \, i \bar{\psi} \gamma_{\nu} D^{\nu} \psi
 = \int d^4 x \, i \Psi^{\dag} \sigma_{\nu} D^{\nu} \Psi.
\label{a7}
\end{equation}
This is invariant
under the $SU(2N_f)$ transformation of $\Psi$ given as
\begin{align}
 \Psi \rightarrow g \Psi, \quad
 (g \in SU(2N_f)).
\label{a7a}
\end{align}

Similarly, using the field $\Psi$,
we rewrite
$\mathcal{L}_{\textrm{ext-scalar}}$ and 
$\mathcal{L}_{\textrm{ext-vector}}$
in Eqs.~(\ref{a3a}) and (\ref{a3b})
as
\begin{align}
 \mathcal{L}_{\textrm{ext-scalar}}
 =
 \frac{1}{2} \Psi^T \sigma_2 \tau_2 \chi \Psi + (\textrm{h.c}), \quad
 \mathcal{L}_{\textrm{ext-vector}}
 =
 \Psi^{\dag} \sigma_{\mu} G^{\mu} \Psi,
\label{a7b}
\end{align}
where $\chi$ and $G^{\mu}$ are external fields of 
$2N_f \times 2N_f$ matrices
defined as
\begin{align}
 &\chi
 \equiv
 \left(
  \begin{array}{cc}
   \mathcal{Q}-i\mathcal{R} &
   -(\mathcal{S}-i\mathcal{P})^T \\
   \mathcal{S}-i\mathcal{R} &
   (\mathcal{Q} + i\mathcal{R})^{\dag}
  \end{array}
 \right),
\label{a7c}\\
 &G^{\mu}
 \equiv
 \left(
  \begin{array}{cc}
   \mathcal{V}^{\mu} + \mathcal{A}^{\mu} &
   (\mathcal{B}^{\mu} + \mathcal{D}^{\mu})^{\dag} \\
   \mathcal{B}^{\mu} + \mathcal{D}^{\mu} &
   - (\mathcal{V}^{\mu} - \mathcal{A}^{\mu})^T
  \end{array}
 \right).
\label{a7d}
\end{align}
Transformation properties of the external fields under $SU(2N_f)$
are given by
\begin{align}
 G_{\mu} \rightarrow g G_{\mu} g^{\dag} + i g (\partial_{\mu} g^{\dag}),
 \quad
 \chi \rightarrow g^{\ast} \chi g^{\dag}.
\label{a7e}
\end{align}

\section{Explicit realization of the $SU(4)$ generators}
\label{sec:generators}

In this appendix, we show the explicit representation
of the generators of $SU(4)$.
We consider the form of the generators following Ref.~\cite{LSS}
for convenience.
They can be represented as
\begin{equation}
 S^{\alpha} = \frac{1}{2 \sqrt{2}}
 \left(
  \begin{array}{cc}
   \mbf{A} & \mbf{B} \\
   \mbf{B}^{\dag} & - \mbf{A}^T
  \end{array}
 \right), \quad
 X^a = \frac{1}{2 \sqrt{2}}
 \left(
  \begin{array}{cc}
   \mbf{C} & \mbf{D} \\
   \mbf{D}^{\dag} & \mbf{C}^T
  \end{array}
 \right),
\label{b1}
\end{equation}
where $\mbf{A}$ is hermitian, $\mbf{C}$ is hermitian and traceless,
$\mbf{B}^T = \mbf{B}$ and $\mbf{D}^T = -\mbf{D}$.
The $\{ S \}$ are also $Sp(4)$ generators
since they obey Eq.~(\ref{21}). We define
\begin{equation}
 S^{\alpha} = \frac{1}{2 \sqrt{2}}
 \left(
  \begin{array}{cc}
   \tau^{\alpha} & \mbf{0} \\
   \mbf{0}       & -(\tau^{\alpha})^T
  \end{array}
 \right), \quad (\, \alpha=1,2,3,4 \,).
\label{b2}
\end{equation}
For $\alpha=1,2,3$, we have the standard Pauli matrices,
while for $\alpha = 4$ we define $\tau^4 = \mbf{1}$.
These are simply the generators for $SU(2) \times U(1)$.
For $\alpha=5,...,10$
\begin{equation}
 S^{\alpha} = \frac{1}{2 \sqrt{2}}
 \left(
  \begin{array}{cc}
   \mbf{0}             & B^{\alpha} \\
   (B^{\alpha})^{\dag} & \mbf{0}
  \end{array}
 \right), \quad (\alpha=5,...,10),
\label{b3}
\end{equation}
and
\begin{equation}
 \, B^5 = \mbf{1}_2, \, \, B^6 = i\mbf{1}_2, \, \, B^7 = \tau^3, \, \, 
 B^8 = i\tau^3, \, \, B^9 = \tau^1, \, \, B^{10} = i\tau^1.
\label{b4}
\end{equation}
The five broken generators $\{ X \}$ are
\begin{equation}
 X^a = \frac{1}{2 \sqrt{2}}
 \left(
  \begin{array}{cc}
   \tau^a  & \mbf{0} \\
   \mbf{0}   & (\tau^a)^T
  \end{array}
 \right), \quad (a=1,2,3),
\label{b5}
\end{equation}
where $\tau^a$ are the standard Pauli matrices. For $a=4,5$
\begin{equation}
 X^a = \frac{1}{2 \sqrt{2}}
 \left(
  \begin{array}{cc}
   \mbf{0}      & D^a \\
   (D^a)^{\dag} & \mbf{0}
  \end{array}
 \right), \quad (a=4,5),
\label{b6}
\end{equation}
and
\begin{equation}
 D^4 = \tau^2, \, \, D^5 = i\tau^2.
\label{b7}
\end{equation}
The generators are normalized as follows:
\begin{align}
 \tr( S^{\alpha} S^{\beta} ) = \frac{1}{2} \delta^{\alpha \beta}, \quad
 \tr( X^a X^b ) = \frac{1}{2} \delta^{ab}, \quad
 \tr( S^{\alpha} X^a) = 0.
\label{b8}
\end{align}

\section{$\mathcal{O}(p^4)$ HLS Lagrangian}
\label{sec:O(p4)-lag}

In this appendix, we present a complete list
of the $\mathcal{O} (p^4)$ HLS Lagrangian for general $N_f$ and
$N_C=2$, following
Refs.~\cite{HY-PR,Tanab}.
For the construction, we need the building blocks
\begin{align}
 \hat{\mathcal{V}}_{\mu \nu}
 =
 \frac{1}{2}
 [
  \xi_R \bar{\Sigma} G^T_{\mu \nu} \bar{\Sigma} \xi^{\dag}_R +
  \xi_L G_{\mu \nu} \xi^{\dag}_L
 ], \quad
 \hat{\mathcal{A}}_{\mu \nu}
 =
 \frac{1}{2}
 [
  \xi_R \bar{\Sigma} G^T_{\mu \nu} \bar{\Sigma} \xi^{\dag}_R -
  \xi_L G_{\mu \nu} \xi^{\dag}_L
 ],
\label{f5}
\end{align}
where $G_{\mu\nu}$ is the filed strength of the
external chiral gauge fields defined as
\begin{align}
 G_{\mu \nu} \equiv
 \partial_{\mu} G_{\nu} - \partial_{\nu} G_{\mu}
 -i [ G_{\mu}, G_{\nu} ].
\label{f5a}
\end{align}
From Eq.~(\ref{f5}) together with other building blocks
in TABLE~\ref{tab:trance},
a complete list of the $\mathcal{O} (p^4)$ Lagrangian
invariant under the $C$ and $P$ transformation 
is constructed as
\begin{align}
 \mathcal{L}_{(4)}
 =
 \mathcal{L}_{(4)y} + \mathcal{L}_{(4)w} + \mathcal{L}_{(4)z},
\label{f1}
\end{align}
where
\begin{align}
 \mathcal{L}_{(4)y}
 &=
 y_1
 \tr[ 
     \hat{\alpha}_{\perp \mu} \hat{\alpha}^{\mu}_{\perp}
     \hat{\alpha}_{\perp \nu} \hat{\alpha}^{\nu}_{\perp}
    ]
 + y_2
 \tr[ 
     \hat{\alpha}_{\perp \mu} \hat{\alpha}_{\perp \nu}
     \hat{\alpha}^{\mu}_{\perp} \hat{\alpha}^{\nu}_{\perp}
    ]
 \nonumber\\
 &\quad +
 y_3
 \tr[ 
     \hat{\alpha}_{\parallel \mu} \hat{\alpha}^{\mu}_{\parallel}
     \hat{\alpha}_{\parallel \nu} \hat{\alpha}^{\nu}_{\parallel}
    ]
 + y_4
 \tr[ 
     \hat{\alpha}_{\parallel \mu} \hat{\alpha}_{\parallel \nu}
     \hat{\alpha}^{\mu}_{\parallel} \hat{\alpha}^{\nu}_{\parallel}
    ]
 \nonumber\\
 &\quad +
 y_5
 \tr[ 
     \hat{\alpha}_{\perp \mu} \hat{\alpha}^{\mu}_{\perp}
     \hat{\alpha}_{\parallel \nu} \hat{\alpha}^{\nu}_{\parallel}
    ]
 + y_6
 \tr[ 
     \hat{\alpha}_{\perp \mu} \hat{\alpha}_{\perp \nu}
     \hat{\alpha}^{\mu}_{\parallel} \hat{\alpha}^{\nu}_{\parallel}
    ]
 + y_7
 \tr[ 
     \hat{\alpha}_{\perp \mu} \hat{\alpha}_{\perp \nu}
     \hat{\alpha}^{\nu}_{\parallel} \hat{\alpha}^{\mu}_{\parallel}
    ]
 \nonumber\\
 &\quad +
 y_8
 \{
   \tr[ 
       \hat{\alpha}_{\perp \mu} \hat{\alpha}^{\mu}_{\parallel}
       \hat{\alpha}_{\perp \nu} \hat{\alpha}^{\nu}_{\parallel}
      ]
   +
   \tr[ 
       \hat{\alpha}_{\perp \mu} \hat{\alpha}_{\parallel \nu}
       \hat{\alpha}^{\nu}_{\perp} \hat{\alpha}^{\mu}_{\parallel}
      ]
  \}
 + y_9
 \tr[ 
     \hat{\alpha}_{\perp \mu} \hat{\alpha}_{\parallel \nu}
     \hat{\alpha}^{\mu}_{\perp} \hat{\alpha}^{\nu}_{\parallel}
    ]
 \nonumber\\
 &\quad +
 y_{10}
 ( \tr[ \hat{\alpha}_{\perp \mu} \hat{\alpha}^{\mu}_{\perp} ])^2
 + y_{11}
 \tr[ \hat{\alpha}_{\perp \mu} \hat{\alpha}_{\perp \nu} ]
 \tr[ \hat{\alpha}^{\mu}_{\perp} \hat{\alpha}^{\nu}_{\perp} ]
 \nonumber\\
 &\quad +
 y_{12}
 ( \tr[ \hat{\alpha}_{\parallel \mu} \hat{\alpha}^{\mu}_{\parallel} ])^2
 + y_{13}
 \tr[ \hat{\alpha}_{\parallel \mu} \hat{\alpha}_{\parallel \nu} ]
 \tr[ \hat{\alpha}^{\mu}_{\parallel} \hat{\alpha}^{\nu}_{\parallel} ]
 \nonumber\\
 &\quad +
 y_{14}
 \tr[ \hat{\alpha}_{\perp \mu} \hat{\alpha}^{\mu}_{\perp} ]
 \tr[ \hat{\alpha}_{\parallel \nu} \hat{\alpha}^{\nu}_{\parallel} ]
 + y_{15}
 \tr[ \hat{\alpha}_{\perp \mu} \hat{\alpha}_{\perp \nu} ]
 \tr[ \hat{\alpha}^{\mu}_{\parallel} \hat{\alpha}^{\nu}_{\parallel} ],
\label{f2}\\
 \mathcal{L}_{(4)w}
 &=
 w_1 \frac{f^2_{\chi}}{f^2_{\pi}}
 \tr[ 
      \hat{\alpha}_{\perp \mu} \hat{\alpha}^{\mu}_{\perp}
      (\hat{\chi} + \hat{\chi}^{\dag})
    ]
 + w_2 \frac{f^2_{\chi}}{f^2_{\pi}}
 \tr[ \hat{\alpha}_{\perp \mu} \hat{\alpha}^{\mu}_{\perp} ]
 \tr[ \hat{\chi} + \hat{\chi}^{\dag} ]
 \nonumber\\
 &\quad +
 w_3 \frac{f^2_{\chi}}{f^2_{\pi}}
 \tr[ 
      \hat{\alpha}_{\parallel \mu} \hat{\alpha}^{\mu}_{\parallel}
      (\hat{\chi} + \hat{\chi}^{\dag})
    ]
 + w_4 \frac{f^2_{\chi}}{f^2_{\pi}}
 \tr[ \hat{\alpha}_{\parallel \mu} \hat{\alpha}^{\mu}_{\parallel} ]
 \tr[ \hat{\chi} + \hat{\chi}^{\dag} ]
 \nonumber\\
 &\quad +
 w_5 \frac{f^2_{\chi}}{f^2_{\pi}}
 \tr[
     (
      \hat{\alpha}^{\mu}_{\parallel} \hat{\alpha}_{\perp \mu} -
      \hat{\alpha}_{\perp \mu} \hat{\alpha}^{\mu}_{\parallel}
     )
     (\hat{\chi} - \hat{\chi}^{\dag})
    ]
 \nonumber\\
 &\quad +
 w_6 \frac{f^2_{\chi}}{f^2_{\pi}}
 \tr[ (\hat{\chi} + \hat{\chi}^{\dag})^2 ]
 + w_7
 \frac{f^2_{\chi}}{f^2_{\pi}}
 ( \tr[ \hat{\chi} + \hat{\chi}^{\dag} ] )^2
 \nonumber\\
 &\quad +
 w_8 \frac{f^2_{\chi}}{f^2_{\pi}}
 \tr[ (\hat{\chi} - \hat{\chi}^{\dag})^2 ]
 + w_9
 \frac{f^2_{\chi}}{f^2_{\pi}}
 ( \tr[ \hat{\chi} - \hat{\chi}^{\dag} ] )^2,
\label{f3}\\
 \mathcal{L}_{(4)z}
 &=
 z_1 \tr[ \hat{\mathcal{V}}_{\mu \nu} \hat{\mathcal{V}}^{\mu \nu} ]
 + z_2 \tr[ \hat{\mathcal{A}}_{\mu \nu} \hat{\mathcal{A}}^{\mu \nu} ]
 + z_3 \tr[ \hat{\mathcal{V}}_{\mu \nu} V^{\mu \nu} ]
 \nonumber\\
 &\quad +
 i z_4
 \tr[
     V_{\mu \nu}
     \hat{\alpha}^{\mu}_{\perp} \hat{\alpha}^{\nu}_{\perp}
    ]
 + i z_5
 \tr[
     V_{\mu \nu}
     \hat{\alpha}^{\mu}_{\parallel} \hat{\alpha}^{\nu}_{\parallel}
    ]
 \nonumber\\
 &\quad +
 i z_6
 \tr[
     \hat{\mathcal{V}}_{\mu \nu}
     \hat{\alpha}^{\mu}_{\perp} \hat{\alpha}^{\nu}_{\perp}
    ]
 + i z_7
 \tr[
     \hat{\mathcal{V}}_{\mu \nu}
     \hat{\alpha}^{\mu}_{\parallel} \hat{\alpha}^{\nu}_{\parallel}
    ]
 - i z_8
 \tr[
     \hat{\mathcal{A}}_{\mu \nu}
     (
      \hat{\alpha}^{\mu}_{\perp} \hat{\alpha}^{\nu}_{\parallel} +
      \hat{\alpha}^{\mu}_{\parallel} \hat{\alpha}^{\nu}_{\perp}
     )
    ].
\label{f4}
\end{align}


\end{document}